\documentstyle[twoside,fleqn,espcrc2,epsf]{article}

\newcommand{\AmS}{{\protect\the\textfont2
  A\kern-.1667em\lower.5ex\hbox{M}\kern-.125emS}}

\title{\vspace{-3.65cm}
       {\normalsize DESY 96--003}    \\[-0.2cm]
       {\normalsize HUB--EP--96/1}   \\[-0.2cm]
       {\normalsize HLRZ 2/96}       \\[-0.2cm]
       {\normalsize January 1996}    \\
       \vspace{1.25cm}
       The Status of Lattice Calculations of the Nucleon Structure Functions%
            \thanks{Talk presented by R. Horsley at the
                    29th Symposium on the
                    Theory of Elementary Particles, Buckow, Germany.}}

\author{M. G\"ockeler%
           \address{H{\"o}chstleistungsrechenzentrum HLRZ,
                    c/o Forschungszentrum J{\"u}lich, D-52425 J{\"u}lich,
                                                             Germany},
        R. Horsley%
           \address{Institut f\"ur Physik, Humboldt-Universit\"at,
                    D-10115 Berlin, Germany},
        E.-M. Ilgenfritz$^{\rm b}$,
        H. Oelrich$^{\rm a}$,
        H. Perlt%
           \address{Fak. f. Physik und Geowiss., Universit\"at Leipzig,
                    Augustusplatz 10--11, D-04109 Leipzig, Germany},
        P. Rakow$^{\rm a}$,
        G. Schierholz$^{\rm a,}$ \hspace{-0.27cm}
           \address{Deutsches Elektronen-Synchrotron DESY,
                    Notkestra{\ss}e 85, D-22603 Hamburg, Germany}
        and
        A.~Schiller$^{\rm c}$}

\begin{document}

\begin{abstract}
We review our progress on the lattice calculation of low moments
of both the unpolarised and polarised nucleon structure functions.
\end{abstract}

\maketitle

\setcounter{footnote}{0}

\section{INTRODUCTION}
\label{intro}

Much of our knowledge about QCD and the structure of nucleons
has been derived from charged current Deep Inelastic Scattering (DIS)
experiments in which a lepton (usually an electron or muon) radiates
a virtual photon at large spacelike momentum $-q^2 = Q^2$.
This interacts with a nucleon (usually a proton)
and measuring the total cross section yields information on
$4$ structure functions for the nucleon -- $F_1$, $F_2$
when polarisations are summed over and $g_1$, $g_2$
when the incident lepton beam and the proton have definite
polarisations. The structure functions are functions of the Bjorken
variable $x$ ($0 \leq x \leq 1$) and $Q^2$. While the unpolarised case has
been studied experimentally for many years (since the pioneering discoveries
at SLAC), only recently have experiments been reported
with polarised beams and targets (\cite{ashman89a,abe95a}).

Theoretically, deviations from the simplest parton model (in which the
hadron may be considered as a collection of non-interacting quarks) is taken
as strong evidence for the evidence of QCD, where we have interacting
quarks and gluons. A direct theoretical calculation of the structure
functions seems not to be possible; however using the Wilson Operator Product
Expansion (OPE) we may relate moments of the structure functions
to matrix elements of certain operators in a twist or Taylor
expansion in $1/Q^2$. For the unpolarised structure functions we find%
\footnote{We use
$\langle \vec{p},\vec{s}| \vec{p}^\prime,\vec{s}^\prime \rangle
= (2\pi)^3 2 E_{\vec{p}} \delta({\vec{p}-\vec{p}^\prime})
\delta_{\vec{s},\vec{s}\prime}$ with $s^2 = -m^2$. The Wilson
coefficients $E \equiv E(\mu^2/Q^2,g(\mu))$ are known
perturbatively, for example for the charged current
we have $E^{(q)}_{n} = e^{(q)2}(1+ O(g^2))$,
$E^{(g)}_{n} = \sum_q e^{(q)2}O(g^2)$.},
\begin{eqnarray}
 \lefteqn{2\int_0^1 \! dx x^{n-1} F_1(x,Q^2)}   \nonumber \\ 
   &=& \sum_f E^{(f)}_{F_1,n} v_n^{(f)}(\mu) + O(1/Q^2),
                                                \nonumber \\
 \lefteqn{\int_0^1 \! dx x^{n-2} F_2(x,Q^2)}    \nonumber \\
   &=& \sum_f E^{(f)}_{F_2,n} v_n^{(f)}(\mu) + O(1/Q^2),
\label{intro.1}
\end{eqnarray}
where $n$ is even starting at $2$, $f = u, d, s, g$ and
\begin{equation}
 \langle \vec{p},\vec{s}| {\cal O}^{ \{ \mu_1\cdots\mu_n \} }_f
                                     | \vec{p},\vec{s} \rangle
   = 2 v^{(f)}_n  [p^{\mu_1} \cdots p^{\mu_n} - \mbox{\rm Tr}],
\label{intro.2}
\end{equation}
with
\begin{eqnarray}
 {\cal O}^{\mu_1\cdots\mu_n}_q
   &=& \left(\frac{i}{2}\right)^{n-1}\bar{q}\gamma^{\mu_1}
           \stackrel{\leftrightarrow}{D}^{\mu_2} \cdots
           \stackrel{\leftrightarrow}{D}^{\mu_n}q,             \\
 {\cal O}^{\mu_1\cdots\mu_n}_g
   &=& i^{n-2}
          \mbox{\rm{Tr}} F^{\mu_1\alpha}
                D^{\mu_2} \cdots D^{\mu_{n-1}}
                F^{\phantom{\alpha} \mu_n}_{\alpha}. \nonumber
\label{intro.3}
\end{eqnarray}
For the polarised structure functions,
\begin{eqnarray}
 \lefteqn{2\int_0^1 \! dx x^n g_1(x,Q^2)}              \nonumber \\ 
   &=& {1\over 2} \sum_f E^{(f)}_{g_1,n}
                  a_n^{(f)}(\mu) + O(1/Q^2),           \nonumber \\
 \lefteqn{2\int_0^1 \! dx x^n g_2(x,Q^2)}              \nonumber \\
   &=& {1\over 2} {n\over{n+1}} \sum_f (
       E^{(f)}_{g_2,n} d_n^{(f)}(\mu) -
       E^{(f)}_{g_1,n} a_n^{(f)}(\mu) )                \nonumber \\
   & & + O(1/Q^2),
\label{intro.4}
\end{eqnarray}
and
\begin{eqnarray}
 \lefteqn{\langle \vec{p},\vec{s}| 
 {\cal O}_{5f}^{ \{ \sigma\mu_1\cdots\mu_n \} }
                           | \vec{p},\vec{s} \rangle} \nonumber \\
   &=& {a^{(f)}_n\over{n+1}} [ s^\sigma p^{\mu_1} \cdots p^{\mu_n} + \cdots ],
                                              \nonumber \\
 \lefteqn{\langle \vec{p},\vec{s}| 
 {\cal O}_{5f}^{ [  \sigma \{ \mu_1 ] \cdots
                                  \mu_n \} }
                           | \vec{p},\vec{s} \rangle} \nonumber \\
   &=& {d^{(f)}_n\over{n+1}} [ (s^\sigma p^{\mu_1} - s^{\mu_1} p^\sigma)
                         p^{\mu_2}\cdots p^{\mu_n} + \cdots ],
\label{intro.5}
\end{eqnarray}
with
\begin{eqnarray}
 {\cal O}_{5q}^{\sigma\mu_1\cdots\mu_n}
    &=& \left(\frac{i}{2}\right)^n\bar{q}\gamma^{\sigma} \gamma_5
        \stackrel{\leftrightarrow}{D}^{\mu_1} \cdots
        \stackrel{\leftrightarrow}{D}^{\mu_n}q, \\
 {\cal O}_{5g}^{\sigma\mu_1\cdots\mu_n}
    &=& i^{n-1}
        \mbox{\rm{Tr}} \tilde{F}^{\sigma\alpha}
        D^{\mu_1} \cdots D^{\mu_{n-1}}
        F^{\phantom{\alpha} \mu_n}_{\alpha}.     \nonumber
\label{intro.6}
\end{eqnarray}
For $g_1$ we start with $n=0$, while for $g_2$ we begin with $n=2$
(${\cal O}_{5g}, n=0$ is a special case).

While the Wilson coefficients, $E$, can be calculated perturbatively,
the matrix elements cannot: a non-perturbative method
must be employed. In principle lattice gauge theories provide
such a method to compute these matrix elements from first principles.
In this talk we shall describe our progress towards this goal,
\cite{goeckeler94a,goeckeler95a}, and
try to discuss the problems that must be overcome.
We will not dwell on lattice details and only give results
in the form of Edinburgh (or APE) graphs which are plotted
in terms of physical quantities.

The moments, eq.\,(\ref{intro.1}), have a parton model
interpretation, being powers of the fraction of the nucleon
momentum carried by the parton
\begin{eqnarray}
   v_n^{(q)}(\mu)
     &=& \langle x^{n-1} \rangle^{(q)}(\mu)      \\
     &=& \! \int_0^1 \! dx x^{n-1} [ q(x,\mu) + (-1)^n \bar{q}(x,\mu) ],
                                                 \nonumber
\label{intro.7}
\end{eqnarray}
where $q(x,\mu) = q_\uparrow(x,\mu) + q_\downarrow(x,\mu)$ (and similarly
for $\bar{q}(x,\mu)$) with $q_\updownarrow(x,\mu)$ being the quark
distributions at some scale $\mu$ and spin $\updownarrow$.
For the gluon
\begin{eqnarray}
   v_n^{(g)}(\mu)
     &=& \langle x^{n-1} \rangle^{(g)}(\mu)   \nonumber \\
     &=& \! \int_0^1 \! dx x^{n-1} g(x,\mu).
\label{intro.8}
\end{eqnarray}
Often we re-write the distributions in terms of valence, $v$ and sea,
$S$ contributions ($q = u, d$),
\begin{eqnarray}
   q(x,\mu)        &\approx& q_v(x,\mu) + S(x,\mu), \nonumber \\
   \bar{q}(x,\mu)  &\approx& S(x,\mu),              \nonumber \\ 
   s(x,\mu)        &\approx& S(x,\mu),
\label{intro.9}
\end{eqnarray}
assuming $SU_F(3)$ symmetry. While in $e(\mu) N$ DIS we can use only the
charge conjugation positive moments (ie for even $n$), in $\nu N$ experiments
we do not have this restriction and can use all moments.
The lowest moment is particularly interesting, as due to conservation
of the energy momentum tensor we have the sum rule
$\sum_q \langle x \rangle^{(q)} + \langle x \rangle^{(g)} = 1$.

For the polarised structure functions a similar interpretation
holds for $g_1$,
\begin{eqnarray}
   a_n^{(q)}(\mu)
\hspace{-0.10in}  &=&  \hspace{-0.10in}
   2 \langle x^n \rangle^{(q)}_5(\mu) \\
\hspace{-0.10in}  &=&  \hspace{-0.10in}
   2 \int_0^1 \! dx x^n
                    [ \Delta q(x,\mu) + (-1)^n \Delta \bar{q}(x,\mu) ],
                                                      \nonumber
\label{intro.10}
\end{eqnarray}
where $\Delta q(x,\mu) = q_\uparrow(x,\mu) - q_\downarrow(x,\mu)$
(and similarly for $\Delta \bar{q}(x,\mu)$). The suitably modified
eq.\,(\ref{intro.9}) is also taken to hold.
Again the lowest moment is of particular interest because it can be related
to the fraction of the spin carried by the quarks in the nucleon. 
Conventionally we define $\Delta q(\mu)$ by
\begin{equation}
   \Delta q(\mu) = \int_0^1 \! dx
                      [ \Delta q(x,\mu) + \Delta \bar{q}(x,\mu) ],
\label{intro.11}
\end{equation}
and then
\begin{equation}
   2\Delta q = a_0^{(q)}.
\label{intro.12}
\end{equation}
$g_2$ contains not only $a_n$ (the so-called Wandzura-Wilczek
contribution to $g_2$) but also $d_n$ -- a 
twist-$3$ contribution. It has been argued that 
the $n=0$ moment of $g_2$ should be $0$, but it is not
possible to check this sum rule on a lattice. Finally we note that
$g_2$ does not have a partonic interpretation.

\section{THE APE PLOT}
\label{lattice}

We shall now motivate the idea of an Edinburgh (or APE) plot. In a
lattice calculation, we must first Euclideanise ($t \to -it$) and then
discretise the QCD action and operators. This introduces a lattice spacing
$a$ as an ultraviolet cutoff in addition to the (bare) coupling constant
$\beta = 6/g^2$ and for the Wilson formulation of lattice fermions,
$1/\kappa$, which is related to the quark mass. (We shall
only consider $m_u = m_d$ here.)

We first consider measuring the light hadron spectrum.
Numerically we measure dimensionless numbers representing the different
masses. In a region where scaling is taking place, \cite{kenway93a}, then
\begin{eqnarray}
  {m_h^{phys} \over m_{h'}^{phys}}
  \hspace{-0.1in} &=& \hspace{-0.1in} 
     \left. {m_h \over m_{h'}}
     \right|_{\mbox{\small{quark mass}} \equiv m_\pi/m_N, m_K/m_N, \ldots}
                                      \nonumber \\
     &\stackrel{g^2 \to 0}{\to}& \mbox{const.}
\label{lattice.1}
\end{eqnarray}
Thus if we plot, for example $m_N/m_\rho$ as a function of $m_\pi/m_\rho$
(here we will take $m_N/m_\rho = f( (m_\pi/m_\rho)^2 ))$, we must
be in a region where we see universal scaling, ie for different 
$g$ values the results lie on the same curve. The lattice spacing can
be found from $a =  {m_h / m_{h}^{phys}}$.

\begin{figure}[hbt]
\vspace*{-2.25cm}
\hspace*{-1.50cm}
\epsfxsize=10.0cm \epsfbox{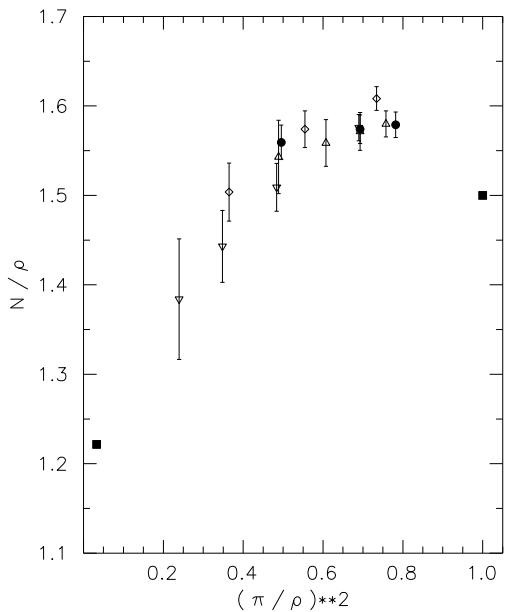}  
\vspace*{-4.50cm}
\caption{\footnotesize The APE plot $m_N/m_\rho$ against $(m_\pi/m_\rho)^2$
         for our results (filled circles,
         $\beta = 6.0, 16^3\times 32$ lattice),
         [4].
         For comparison we also show the results of
         [6], (triangles, $\beta = 6.0, 18^3\times 32$ lattice),
         [7] (inverted triangles,
         $\beta = 6.0, 24^3\times 32$ lattice), and [8]
         (diamonds, $\beta = 5.7, 16^3\times 20$ lattice). The filled
         squares are the experimental result and
         heavy quark limit $(1,3/2)$.}
\vspace*{-0.75cm}
\label{mass_ape}
\end{figure}

Measuring the various masses yields the plot shown in Fig.~\ref{mass_ape}
The plot allows the whole range of quark mass to be shown, from the
heavy quark mass limit to the physical point. In fact from
the lattice many quark mass worlds can be computed. (Unfortunately
everything except the real world: this is because
the $u$ and $d$ quark masses are very close to the chiral limit,
which numerically is difficult to realise.) Our lightest
quark mass lies at about the mass of the strange quark (in the middle
of the plot, $(m_\pi/m_\rho)^2 \approx 0.5$). Do we have universal scaling?
Roughly the points seem to lie in a band -- a good sign, although the
smaller coupling values seem to lie slightly above the larger coupling value.
Are there finite size effects? Looking at the lightest quark masses
we see a slight dropping of the value
when increasing the spatial lattice size. This might be interpreted
as a sign of some finite volume effects. We note also  the difficulty
of linear extrapolations -- using our results alone
would give a rather different result to the physical result.
Using lighter quark mass results as well gives a more satisfactory
extrapolation, \cite{goeckeler95a}. The curve from the heavy
quark mass world to the physical world is certainly rather
unpleasant -- the curve first goes up before sinking down.
We have also used the quenched approximation in which the
fermion determinant is ignored (otherwise the computations are too costly).
This introduces an uncontrolled approximation -- for example it is
thus not clear that the ratio in eq.\,(\ref{lattice.1}) need be
the physical value. Finally as we have a choice of which
hadron mass to use to find $a$, if we
are not in the neighbourhood of the physical point, we will get
ambiguities in the value. From \cite{goeckeler95a}%
\footnote{$m_N = 0.440(4)$, $m_\rho = 0.321(4)$ at $\kappa_c = 0.15699(5)$.},
we find $a \approx 2.0-2.5 \mbox{GeV}^{-1}$ (using $m_N$, $m_\rho$).

\section{LOW MOMENTS}
\label{low}

We shall now present our results for the low moments of the structure
functions. The computational method is to evaluate the nucleon
two point functions with a bilinear quark or gluon operator insertion,
leading to a numerical estimate of the matrix elements in
eqs.\,(\ref{intro.2},\ref{intro.5}). There are two basic types:
The first is quark insertion in one of the nucleon quark lines
(quark line connected). As this involves a valence quark, we shall
take this as probing the valence distributions in eq.\,(\ref{intro.9}).
In the second type the operator interacts via the exchange of gluons
with the nucleon (for quark line disconnected and gluon operators).
These are taken as probes of the sea and gluon
distributions. (Of course as we work in the quenched approximation
there may be some distortions in the various distributions.) 
Due to technical difficulties, such as the
operator growing too large in comparison with the lattice spatial size
and the mixing of the lattice operators with lower dimensional
operators, it has only been possible to measure the lowest
three quark moments for the unpolarised valence structure functions,
to give an estimate for the lowest gluon moment
and for the polarised valence structure functions to compute
$a_0$, $a_2$ and $d_2$. (Some sea quark distributions also have been
attempted in \cite{fukugita94a,dong95a}.) After computation
the lattice matrix elements must be renormalised.
For this we use at present the one--loop perturbation
results, \cite{goeckeler95a}. (However improvements are being developed,
\cite{martinelli95a}.)
Finally we shall quote all our results at the reference scale,
\begin{equation}
   Q^2 = \mu^2 = {1\over a^2} \approx 4-6 \mbox{GeV}^2
\label{low.1}
\end{equation}
(to avoid having to re-sum large logarithms in $Z$ or the
Wilson coefficients).

\subsection*{Unpolarised}
\label{unpolarised}

\begin{figure}[hbt]
\vspace*{-1.75cm}
\hspace*{-1.50cm}
\epsfxsize=10.0cm \epsfbox{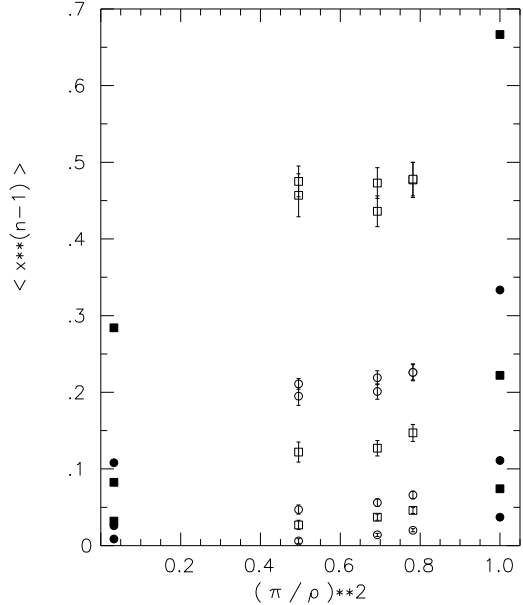}  
\vspace*{-4.50cm}
\caption{\footnotesize The APE plot for valence results:
         $\langle x^{n-1} \rangle^{(q)}$ with $n = 2,3,4$,
         top to bottom, in turn first for the $u$ quarks (squares)
         and then for the $d$ quarks (circles). (For $\langle x \rangle$,
         we have evaluated the matrix elements in two
         distinct ways, [4], methods $a, b$.) 
         The experimental numbers and heavy quark limit are the
         filled symbols.}
\vspace*{-0.75cm}
\label{xall}
\end{figure}

\begin{figure}[hbt]
\vspace*{-1.75cm}
\hspace*{-1.50cm}
\epsfxsize=10.0cm \epsfbox{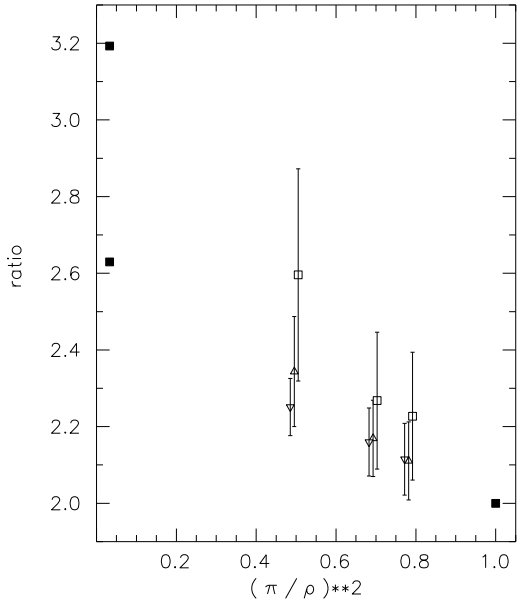}  
\vspace*{-4.50cm}
\caption{\footnotesize The APE plot for the ratio of the valence
         moments:
         $\langle x^{n-1}\rangle^{(u)}/\langle x^{n-1}\rangle^{(d)}$.
         Method $a$ ($n=2$) are triangles.
         Method $b$ (inverted triangles) and $n=3$ (squares) are slightly
         displaced for clarity.}
\vspace*{-1.00cm}
\label{xdiv}
\end{figure}

\begin{figure}[hbt]
\vspace*{-1.75cm}
\hspace*{-1.50cm}
\epsfxsize=10.0cm \epsfbox{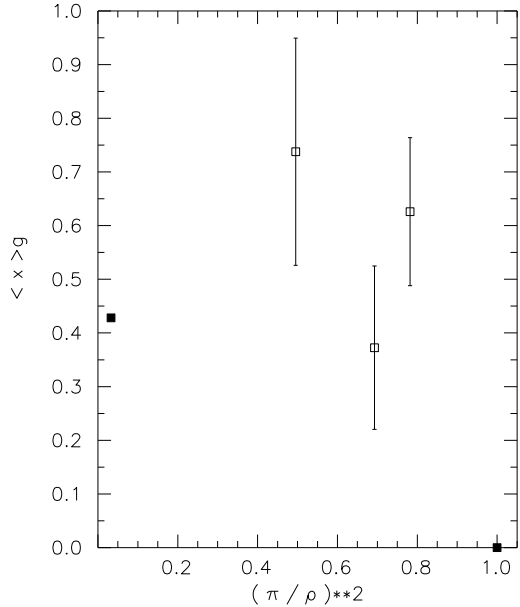}  
\vspace*{-4.50cm}
\caption{\footnotesize The APE plot for $\langle x \rangle^{(g)}$.}
\vspace*{-0.75cm}
\label{x1_gluon}
\end{figure}

In Fig.~\ref{xall} we show the APE plot for the $3$ lowest valence moments.
Also shown are the physical values (taken from \cite{martin93a})
and the heavy quark mass limit%
\footnote{$c_{u_\uparrow}=5/3$, $c_{u_\downarrow}=1/3$,
$c_{d_\uparrow}=1/3$, $c_{d_\downarrow}=2/3$.}
$q_{\updownarrow,v}(x) \to c_{q_\updownarrow} \delta(x-1/3)$,
$S(x) \to 0$.
We see that the results roughly
seem to lie on a smooth extrapolation from the heavy quark mass limit 
to the physical result. (Although the lightest quark mass seems to
be levelling off, this should be checked on a larger lattice to
see that there are no finite volume effects present.) It is to be hoped that
smaller quark masses will continue to drift downwards. Certainly little sign
of overshooting is seen (as happened in the APE mass plot).
In Fig.~\ref{xdiv} we show the ratios of the $u_v$ to $d_v$ moments.
There seem indications of a very smooth (linear?) behaviour with
the quark mass. In Fig.~\ref{x1_gluon} we show the lowest gluon moment.
The experimental result comes from \cite{martin95a}.
(In the heavy quark mass limit we would expect all the nucleon momentum
to reside in the quarks and none in the gluons.) At present the signal is
not so good. This is numerically a difficult calculation however
and requires a large statistic. No statement at present can be made
on the momentum sum rule.

\subsection*{Polarised}
\label{polarised}

We now turn to a consideration of the polarised structure function
moments. We first consider the non-singlet results for $a_0$.
The octet and triplet currents are given by
\begin{eqnarray}
 3F_A - D_A = \Delta u + \Delta d - 2\Delta s
              \approx \Delta u_v + \Delta d_v  \nonumber \\
 F_A + D_A  = g_A =  \Delta u - \Delta d \approx \Delta u_v - \Delta d_v
\label{pol.1}
\end{eqnarray}
and thus we need only look at connected diagrams. Space prevents us
from showing results for $g_A$, $3F_A-D_A$ or $F_A/D_A$, we refer the
interested reader to the nice review \cite{okawa95a}. 
(Comparing the results there with the experimental and the heavy
quark mass limit we again see for $g_A$ an overshoot effect:
we seem to have to climb up to reach the experimental value.)
For the singlet results we have
$\Delta u + \Delta d + \Delta s \equiv \Delta \Sigma$
and so we expect the disconnected terms to be more important.
This is a very difficult problem, \cite{fukugita94a,dong95a}.
Recent numerical estimates however show an encouraging trend to
the experimental result, \cite{okawa95a}.

\begin{figure}[hbt]
\vspace*{-1.75cm}
\hspace*{-1.50cm}
\epsfxsize=10.0cm \epsfbox{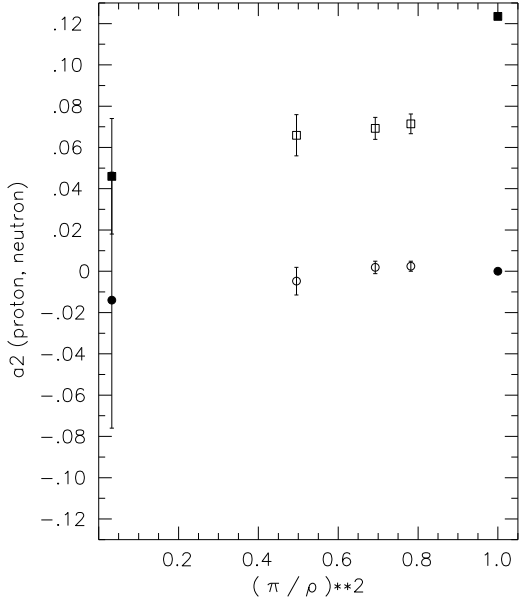}  
\vspace*{-4.5cm}
\caption{\footnotesize The APE plot for $a_2^{p,n}$.
         The squares are the proton results, while the circles
         are the neutron results.}
\vspace*{-0.75cm}
\label{a2pn_ape}
\end{figure}

Of the higher $a$ moments we have been able to look at $a_2$.
Changing to $p=\mbox{proton}$, $n=\mbox{neutron}$ (rather than $u$, $d$),
we plot $a_2$ in Fig.~\ref{a2pn_ape}.
(The numbers for $a_2$ and $d_2$ are conveniently collated in
\cite{mankiewicz95a}.) Again we see a rather 
smooth behaviour between the heavy quark mass limit and the experimental
result. Indeed, perhaps due to the present rather large errors there,
we could claim that we have reasonable agreement with experiment.

\begin{figure}[hbt]
\vspace*{-3.00cm}
\hspace*{-1.50cm}
\epsfxsize=10.0cm \epsfbox{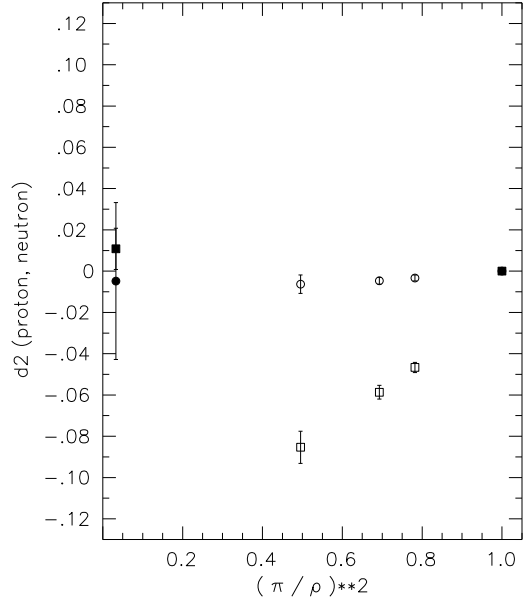}  
\vspace*{-4.5cm}
\caption{\footnotesize The APE plot for  $d_2^{p,n}$. The same
         notation as for Fig.~\ref{a2pn_ape}.}
\vspace*{-0.75cm}
\label{d2pn_ape}
\end{figure}

\begin{figure}[hbt]
\vspace*{-3.00cm}
\hspace*{-1.50cm}
\epsfxsize=10.0cm \epsfbox{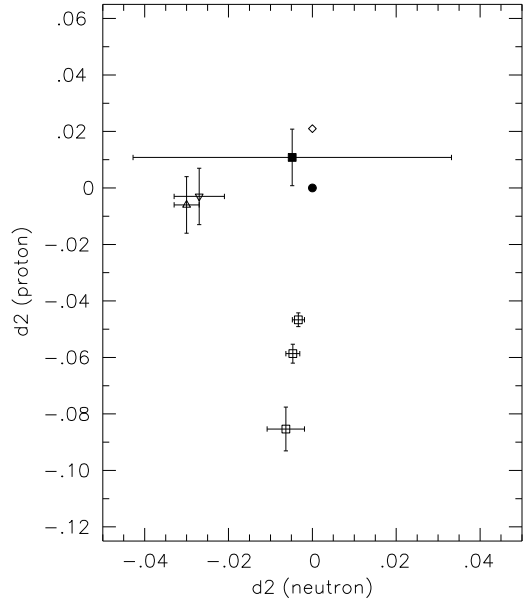}  
\vspace*{-4.5cm}
\caption{\footnotesize $d_2^{(p)}$ versus $d_2^{(n)}$. The lattice results
         are shown as squares, sum rule as (inverted ) triangles, and the bag
         estimate as a diamond. The experimental value is a filled square
         and the heavy quark mass limit is a filled circle.}
\vspace*{-0.75cm}
\label{d2pn_est_ape}
\end{figure}

Finally we turn to an estimation of $d_2$ -- the first non-leading
twist operator in the OPE expansion. In Fig.~\ref{d2pn_ape} we plot our
results. While one might say that for $n$ we are compatible with
the experimental result, for $p$ we have crass disagreement.
At present we have no explanation for this result.
(Indeed as we tend towards the chiral limit, the discrepancy grows.)
In Fig.~\ref{d2pn_est_ape}
we compare our results with other theoretical estimates.
At present the lattice result comes out a poor last. (But one should
note that the most compatible result to the experiment
appears to be the heavy quark mass limit.)

\subsection*{Conclusions}
\label{conclusions}

Although progress has been made, it is clear that much more remains
to be done. In particular for reliable extrapolations to the chiral
limit, smaller quark mass simulations are necessary.

\section*{ACKNOWLEDGMENTS}
\label{acknowledgements}

The numerical calculations were performed on the QUADRICS (Q16 and QH2)
at DESY (Zeuthen) and at Bielefeld University.
We wish to thank both institutions for their support and in particular
the system managers H. Simma, W. Friebel and M. Plagge for their help.

\end{document}